\documentclass{aastex}

\begin{document}

\title{ESTIMATING SUPERMASSIVE BLACK HOLE MASS \\
       THROUGH RADIO/X-RAY LUMINOSITY RELATION\\
       OF X-RAY BRIGHT GALACTIC NUCLEI}

\author{Heon-Young Chang$^1$, Chul-Sung Choi$^2$, and Insu Yi$^1$}

\affil{$^1$ Korea Institute for Advanced Study, Seoul 130-012, Korea\\
$^2$ Korea Astronomy Observatory, Taejon 305-348, Korea}

\email{hyc@kias.re.kr, cschoi@kao.re.kr, iyi@kias.re.kr}

\begin{abstract}

It has been suggested that optically thin and geometrically thick 
accretion flows are responsible for the observed radio/X-ray 
luminosity relation of the X-ray bright galactic nuclei.
 If this is the case then central 
supermassive black hole masses can be estimated directly from 
measurements of the core radio luminosity and the X-ray luminosity, 
provided that properties of such accretion flows are known. Calculated 
ratios of the luminosities are presented in cases of the standard 
ADAF model and modified ADAF models, in which a truncation of inner 
parts of the flows and winds causing a reduction of the infalling 
matter are included. We compare the observed ratio of the luminosities 
with predictions from models of optically thin accretion flows. We also 
discuss the possible effects of the convection in ADAFs. We confirm
that the supermassive black hole (SMBH) mass estimate is possible with 
the radio/X-ray luminosity relation due to ADAF models in the absence
of a radio jet. 
We find that observational data are insufficient to distinguish the
standard ADAF model from its modified models. However, the ADAF model
with convection is inconsistent with  observations, unless microphysics
parameters are to be substantially changed.
High resolution radio observations are required to avoid the contamination 
of other components, such as, a jet component. Otherwise, the SMBH mass 
is inclined to be over-estimated.
\end{abstract}

\keywords{accretion disks -- black hole physics -- galaxies : nuclei -- 
radiation mechanisms : nonthermal -- radio continuum : galaxies -- 
X-rays : galaxies}

\section{INTRODUCTION}

Since the early days of research on quasars and active galactic 
nuclei (AGNs), supermassive black holes (SMBHs) have been considered 
as the most likely power sources of the activity in these objects 
\citep{lyn69, rees84}. 
SMBHs at the centers of all galaxies are now recognised as 
ubiquitous, whose mass  $M_{\rm SMBH}$ is proportional to the spheroidal 
bulge mass of the host galaxy or the galactic bulge luminosity 
\citep{kor95,mag98,rich98}. Evidence for the existence of SMBHs has 
 been found in the center of our Galaxy \citep{eck97,gen97,ghe98} 
and in the weakly active galaxies NGC 1068, NGC 4258 \citep{miyo95}. 
Asymmetric Fe K$\alpha$ emission 
in the X-ray spectra of AGNs (e.g., Tanaka et al. 1995) may show a 
signature of SMBHs, but this remains somewhat speculative as Fe K$\alpha$ 
reverberation signature have not yet been observed (e.g., Reynolds 2000).

Searches for SMBHs are based on spatially resolved 
kinematics.  In the case of AGNs, however, direct detection 
of nuclear SMBHs through stellar-dynamical methods 
has not been achieved due to technical difficulties. The reason is 
that bright AGNs are too bright to resolve 
the light from the surrounding stars and gas, and from the AGN itself 
on arcsecond and smaller angular scales. 
Reverberation mapping \citep{blan82} avoids this problem. In this 
technique, the time delayed response of the emission lines to 
continuum variations is used  to infer a size of the line-emitting 
region. We also have a velocity $V$, which is obtained from 
measurements of the Doppler widths of the variable line components. 
Combining these, a virial mass is estimated by $M_{\rm SMBH} \approx 
V^2 R /G$, where $G$ is the gravitational constant.  
 
On the other hand, several authors have pointed out that reverberation 
mapping yields systemically smaller SMBH masses at a given bulge 
luminosity than do dynamical models of spatially resolved kinematics 
(e.g., Wandel, Perterson, \& Malkan 1999). In the case of normal 
galaxies, the SMBH mass appears to correlate with the galactic bulge 
luminosity, with the SMBH to spheriodal bulge mass ratio $M_{\rm SMBH}
/M_{\rm bulge}=0.006$ \citep{mag98, rich98}. In the case of AGNs, 
however, a significantly lower value of $M_{\rm SMBH}/M_{\rm bulge}$ 
has been found \citep{wan99a}, indicating either a real difference 
between active and inactive galaxies, or that one or both of the 
methods of mass determination is somehow biased. \citet{geb00b}  
 claim that masses derived from reverberation mapping are 
consistent with the recently discovered relation between the SMBH 
mass and the galaxy velocity dispersion derived from 
spatially resolved kinematics. 

Another possible way to infer the presence of SMBHs and to shed light 
on the physical condition is to examine spectral energy distribution
 over a wide range from the radio to the hard X-ray frequencies 
(e.g., Frank et al. 1992; Ho 1999). This emission spectrum is 
produced by an accreting matter, as the surrounding gas accretes 
onto the central SMBH. In such flows the 
core radio luminosity is low and dependent on the mass of the central 
SMBH \citep{ich77,rees82,nayi94,nayi95a,nayi95b,abr95}.
It has been suggested that 
advection-dominated accretion flows (ADAFs) are responsible 
for the observed radio and X-ray luminosities of some of the X-ray bright 
galactic nuclei \citep{fab95,dim97,yib98,yib99}. 
In this case the central SMBH masses can be estimated 
directly from measurements of the X-ray and the core radio luminosities, 
provided that properties of such accretion flows are known. 
We demonstrate that it is relatively effective 
and consequently suitable to estimate SMBH masses of many candidates 
using a method we present in this paper. 

This paper begins with model descriptions for the standard ADAF model 
and modified versions of the model in $\S$  2. We deduce analytical expressions to 
describe the radio/X-ray  luminosities for various models in $\S$ 2. 
We present the calculated ratios of the luminosities and compare them 
with the observational data in $\S$ 3. We discuss possible roles of the convection 
in ADAFs and conclude in $\S$ 4.

\section{RADIO/X-RAY EMISSION FROM ACCRETING SMBHs}

When a mass accretion rate is below a critical rate, the radiative 
cooling is slower than the viscous heating. As a result of this, 
the dissipated accretion energy is inefficiently radiated away 
and  advected inward to the central object with 
the accreted matter (see Narayan and Yi 1995b and references 
therein). The ions are heated by viscous dissipation at a rate of 
$q^+$ per unit volume and the heating is parameterised 
by the viscosity parameter $\alpha$. Since this flow is optically 
thin, the energy transfer from the ions to the electrons is 
inefficient so that the ions reach the virial temperature. The electrons 
cool via synchrotron, bremsstrahlung, and inverse Compton processes. 
Synchrotron radiation is responsible 
for the radio to submillimeter emission, bremsstrahlung emission 
and inverse-Compton scattering are submillimeter to X-ray emission.
Detailed numerical calculations have been performed \citep{na98}, 
and the resulting spectra have been successfully applied to a 
number of extra galactic systems which are supposed to
harbor SMBHs (e.g., Narayan et al. 1995; Lasota et al. 1996; 
Manmoto et al. 1997). 

In this study we apply analytical scaling laws for self-similar 
solutions \citep{nayi94,maha97}. We adopt the following 
dimensionless variables : mass of the SMBH $m=M/M_\odot$; 
radius from the SMBH $r=R/R_g$, where $R_g=2GM/c^2=2.95 
\times 10^5 m~{\rm cm}$; 
and mass accretion rate $\dot{m}=\dot{M}/\dot{M}_{\rm Edd}$,
where $\dot{M}_{\rm Edd}=L_{\rm Edd}/\eta_{\rm eff} c^2= 
1.39 \times 10^{18} m~ {\rm g~ s^{-1}}$ (the Eddington 
accretion rate assuming $\eta_{\rm eff}=0.1)$.
We assume that the flows are spherically symmeric. 
The advection fraction $f$ is determined such that the ion and
electron energy balance equations are met. The quantities of 
interest are the volume-integrated quantities which are obtained 
by integrating  the heating rate and each cooling rates 
throughout the volume of the flows. 
As canonical values in a model for ADAFs parameters are taken to be 
$r_{min}=3$, $r_{max}=10^3$,
$\alpha=0.3$, and $\beta=0.5$ (see, e.g., Narayan \& Yi 1995b). 
In this study, we assume that the electron temperature
is constant for $r< 10^3$, as suggested by \citet{nayi95b}. 
For  given $m, \dot{m}, \alpha, \beta$, 
we may obtain the energy density spectrum with $f$, $T_e$ 
determined as  described below.

\subsection{Emission from 'Standard' ADAFs}

The ions are heated by the viscous dissipation
and the electrons gain energy from the ions
by Coulomb interactions alone. 
We neglect the possibility of heating the electrons, since 
we expect that the fraction of viscous energy transferred 
to the electrons is in the mass ratio of the electron to
the ion, $\sim 1/2000$. 

The synchrotron photons are self-absorbed and give a 
blackbody radiation upto a critical frequency, $\nu_c$. 
For a given $T_e$, the synchrotron spectrum 
$L_{\nu}^{sync}$ is given by
\begin{eqnarray}
L_{\nu}^{sync}&=&s_3(s_1s_2)^{8/5}m^{6/5}\dot{m}^{4/5}
T_e^{21/5}\nu^{2/5}, \label{eq:one}
\end{eqnarray}
where $s_1=1.42 \times 10^9 \alpha^{-1/2}(1-\beta)^{1/2}
c_1^{-1/2}c_3^{1/2}$, $s_2=1.19 \times 10^{-13} x_M$, 
and $s_3=1.05  \times 10^{-24}$, $x_M \equiv 2 \nu/3 \nu_b 
\theta_e^2$, $\nu_b \equiv eB/2 \pi m_e c$ 
\citep{nayi95b,maha96,maha97}.  The radio luminosity 
$L_R$ at $\nu$ is defined by $\nu L_{\nu}$. The highest 
radio frequency arises from the innermost radius of the accretion 
flows, $r_{min} \sim 3$; $\nu_p=s_1 s_2 m^{-1/2}\dot{m}^{1/2}
T_e^{2} r_{min}^{-5/4}$. At this peak frequency the peak radio 
luminosity is given by
\begin{eqnarray}
L_{R}\equiv \nu_p L_{\nu_p}^{sync}=s_3(s_1s_2)^{3}m^{1/2}
\dot{m}^{3/2}T_e^{7}r^{-7/4}_{min}
 \label{eq:two}.
\end{eqnarray}
We obtain the total power due to synchrotron radiation  from
\begin{eqnarray}
P_{sync}&\simeq&\int^{\nu_p}_{0} L_{\nu}^{sync} d\nu =(5/7) 
\nu_p L_{\nu_{p}}^{sync}.
\end{eqnarray}
We set $\nu_{\rm min}=0$, since  $\nu_{\rm min}$  
is much smaller than $\nu_p$. 

Bremsstrahlung emission is due to both electron-electron 
and electron-ion interactions. 
The total bremsstrahlung power is given by
\begin{eqnarray}
P_{brem}&=&4.74 \times 10^{34} \alpha^{-2} c_1^{-2}
\ln(r_{max}/r_{min})F(\theta_e)m\dot{m}^2, 
\label{eq:eight}
\end{eqnarray}
and the spectrum due to bremsstrahlung emission is given by
\begin{eqnarray}
L^{brem}_{\nu}&=&2.29 \times 10^{24} \alpha^{-2} 
c_1^{-2}\ln(r_{max}/r_{min}) F(\theta_e)m\dot{m}^2
T_e^{-1}\exp(-h \nu/k T_e),  \label{eq:three}
\end{eqnarray}
where 
\begin{eqnarray}
F(\theta_e)&=& 4(\frac{2\theta_e}{\pi^3})^{1/2}(1+1.781
\theta_e^{1.34})+1.73\theta_e^{3/2}(1+1.1\theta_e+
\theta_e^2-1.25\theta_e^{5/2}),~~~~~\theta_e<1,
\nonumber \\
&=&(\frac{9\theta_e}{2\pi})[\ln(1.123\theta_e+0.48)+1.5]+
2.3\theta_e(\ln 1.123\theta_e+1.28),~~~~~~~~~~\theta_e>1.
\end{eqnarray}
 
In this study we neglect the Comptonization of bremsstrahlung 
emission, and consider the Comptonization of synchrotron 
emission alone. A contribution by Compton up-scattered 
synchrotron photons to the hard X-ray luminosity becomes 
important as $\dot{m}$ increases, while bremsstrahlung emission 
dominates the hard X-ray luminosity when $\dot{m}$ is substantially low.

We approximate the Comptonized spectrum by assuming that 
all the synchrotron photons to be Comptonized have an 
initial frequency of $\nu_p$. The maximum final frequency 
of the Comptonized photon is $\nu_f=3 k T_e/h$, which 
corresponds to the average energy of the photon for 
saturated Comptonization in the Wien regime. On average, 
we assume that all the photons would see one half the 
total optical depth \citep{maha97}, which is written as
$\tau_{es}=6.2 \alpha^{-1} c_1^{-1} \dot{m} r^{-1/2}$.
The spectrum of the emerging photons at frequency $\nu$ 
has the power-law shape
\begin{eqnarray}
L_{\nu}^{Comp} \simeq L_{\nu_i} (\frac{\nu}{\nu_i})^
{-\alpha_c},  \label{eq:four}
\end{eqnarray}
where $\alpha_c \equiv -\ln \tau_{es}/\ln A$.
The total Compton power is therefore given by
\begin{eqnarray}
P_{comp}&=&\int^{3 k T_e/h}_{\nu_p} L_{\nu}^{Comp} 
d\nu \nonumber \\
&=&\frac{\nu_p L_{\nu_{p}}^{sync}}{1-\alpha_c}
\Biggl\{\Biggl[\frac{6.2 \times 10^{10} T_e}{\nu_p}
\Biggr]^{1-\alpha_c}-1 \Biggr\},
\end{eqnarray}
where $A$ is the mean amplification factor in a single scattering.

For  given $m, \dot{m}, \alpha, \beta$, the total heating 
of the electrons should be  balanced
to the sum of the individual cooling, $Q^{ie}=P_{sync}+
P_{brem}+P_{comp}$. The electron 
temperature is varied until this equality is satisfied. 

\subsection{Emission from Truncated ADAFs and 'Windy' ADAFs} 

ADAFs have the positive Bernoulli constant and, therefore,
are susceptible to outflows \citep{nayi94}. 
It might be the case that radio jets are present near 
the inner region of ADAFs. If so,
the radio luminosity is dominated by jet emission. 
Instead, radio emission due to ADAFs at the high 
frequency will be supressed. In order to simulate this 
situation, we truncate  ADAFs at
a certain inner radius. We suppose that the truncation 
occurs at $25 r_g$. For instance, an observation of NGC 4486 (M87) 
indicates that the jet is formed in a smaller radius than 
$r \approx 30 r_g$ \citep{ju99}. 
This is slso about where is the maximum radius that 15 GHz 
radio emission
could be generated for $m=10^7, \dot{m}=10^{-3}, T_e=10^9$. 
Note that $r=(1/60)^{-4/5} 
\times (\nu/15~ {\rm GHz})^{-4/5}(m/10^7)^{-2/5} 
(\dot{m}/10^{-3})^{2/5} (T_e/10^9)^{8/5}$.
We set $r_{min}$ in equations in the previous subsection to $25 r_g$
 instead of  $3 r_g$ in corresponding equations in this case. 

Since the gas in ADAF solutions is generically unbound, 
winds may  carry away infalling matter
(see, e.g., Blandford \& Begelman 1999; Di Matteo 
et al. 1999; Quataert \& Narayan 1999). 
We allow $\dot{m}$ to vary as $\dot{m}=
\dot{m}_{out}(r/r_{out})^p$  
(see, e.g., Blandford \& Begelman 1999). The mass loss 
causes a significant effect on a model
spectrum for a large value of $p$ \citep{qua99}. 
Bremsstrahlung emission decreases with increasing $p$,
synchrotron emission decreases strongly with increasing $p$ 
due both to the lower density 
 and to the lower $T_e$ near $r \sim 3$. Compton emission 
decreases with increasing $p$
even more strongly than the other two emissions since it 
depends both on $\nu_p L_{\nu_p}$ and
$\alpha_c$. A wind model is incompatible with observation, 
if microphysics parameters we
adopt in the standard ADAF model are more or less correct. 
This conclusion,  however, is inconclusive since there are 
qualitative degeneracies between the wind 
parameter, $p$, and microphysics parameters, $\alpha, 
\beta$, particularly the viscous
heating paramter of the electrons, $\delta$ \citep{qua99}.
As $\dot{m}$ is replaced by $\dot{m}_{out}(r/r_{out})^p$ 
equations for the 'Standard ADAFs'
are to be modified as described below. 
The total heating for ions  $Q^+$ is rewritten as
\begin{eqnarray}
Q^+&=&9.39 \times 10^{38} \frac{1-\beta}{f} c_3 m 
\dot{m}_{out}r^{-p}_{out}\frac{1}{p-1}
 (r^{p-1}_{max}-r^{p-1}_{min}),
\end{eqnarray}
where $p<1$. Similarly, $Q^{ie}$ is given by 
\begin{eqnarray}
Q^{ie}&=&1.2 \times 10^{38} g(\theta_e)\alpha^{-2} 
c_1^{-2}c_3 \beta m \dot{m}^2_{out}
r^{-2p}_{out}\frac{1}{2p-1}(r^{2p-1}_{max}-r^{2p-1}_{min})
\end{eqnarray}
where $p \neq 1/2$.
We  assume that  $x_M$ and $T_e$ are constants with $r$
and  $\dot{m}$ for simplicity
as a first order approximation. Given $x_M$, the cutoff 
frequency at each radius is
given as $\nu_c=s_1 s_2 m^{-1/2}\dot{m}^{1/2}_{out}
r_{out}^{-p/2}T_e^{2} r^{-(2p-5)/4}$. 
The synchrotron spectrum is given  by 
\begin{eqnarray}
L_{\nu}^{sync}&=&s_3(s_1s_2)^{\frac{8}{5-2p}}
m^{\frac{4p-6}{2p-5}}
\dot{m}_{out}^{\frac{4}{5-2p}}r_{out}^{\frac{4p}{2p-5}}
T_e^{\frac{21-2p}{5-2p}}\nu^{\frac{4p-2}{2p-5}}. 
\end{eqnarray}
The peak frequency and the radio luminosity at the peak 
frequency are correspondingly
reduced so that we have
\begin{eqnarray}
\nu_p L_{\nu_p}^{sync}=s_3(s_1s_2)^{3}m^{1/2}
\dot{m}_{out}^{3/2}r_{out}^{-\frac{3}{2}p}
T_e^{7}r^{\frac{6p-7}{4}}_{min}.
\end{eqnarray}
The total synchrotron power is also rewritten by
\begin{eqnarray}
P_{sync}&\simeq&\int^{\nu_p}_{0} L_{\nu}^{sync} d\nu 
=(\frac{2p-5}{6p-7}) \nu_p L_{\nu_{p}}^{sync}.
\end{eqnarray}
For bremsstrahlung emission the total power and the 
spectrum are respectively given by
\begin{eqnarray}
P_{brem}&=&4.74 \times 10^{34} \alpha^{-2} c_1^{-2} 
m \dot{m}_{out}^2 r_{out}^{-2p}
 F(\theta_e)[\frac{1}{2p}(r^{2p}_{max}-r^{2p}_{min})], \\
L^{brem}_{\nu}&=&2.29 \times 10^{24} \alpha^{-2} c_1^{-2} 
m \dot{m}_{out}^2 r_{out}^{-2p}
F(\theta_e)T_e^{-1}\exp(-h \nu/k T_e)\frac{1}{2p}
(r^{2p}_{max}-r^{2p}_{min})]. 
\end{eqnarray}
Note that as the exponent $p$ approaches to 0, $[\frac{1}{2p}
(r^{2p}_{max}-r^{2p}_{min})]$ becomes 
$\ln(r_{max}/r_{min})$. For the Comptonization of 
the synchrotron photons we take $\tau_{es}$
as $\tau_{es}=6.2 \alpha^{-1} c_1^{-1} 
\dot{m}_{out}r^{-p}_{out} 
(r^{(2p-1)/2}_{max}-r^{(2p-1)/2}_{min})$.
The Compton spectrum and the total Compton power are 
respectively given by
\begin{eqnarray}
L_{\nu}^{Comp} &\simeq& L_{\nu_i} 
(\frac{\nu}{\nu_i})^{-\alpha_c}, \\
P_{comp}&=&\frac{\nu_p L_{\nu_{p}}^{sync}}{1-\alpha_c}
\Biggl\{\Biggl[\frac{6.2 \times 10^{10} T_e}
{\nu_p}\Biggr]^{1-\alpha_c}-1 \Biggr\}.
\end{eqnarray}

\section{BLACK HOLE MASSES AND RADIO/X-RAY LUMINOSITIES}

Figure 1 shows a plot of the ratio of the 15 GHz  radio 
luminosity to the 2-10 keV X-ray 
luminosity versus the X-ray luminosity. The solid lines 
are standard ADAF model 
predictions for SMBH masses of $10^6-10^9 M_{\odot}$, the 
dotted lines are ADAF model predictions with 
truncations at $25r_g$. The short dashed lines and 
the long dashed lines are ADAF model predicitons with winds, 
where $p=0.4$ and $p=0.99$, respectively.
When $p=1$, results may represent a case where the convection
is present in ADAFs \citep{qua99}.
For each curve, $\dot{m}$ or $\dot{m}_{out}$ for windy models
varies from $10^{-4}$ (upper left corner) to $10^{-1.6}$ (lower
right corner). The open circles denote the observed core radio 
luminosity at 15 GHz,
and the filled circles the spatially resolved core radio 
luminosity which are converted from
5 GHz to 15 GHz using the $\nu^{7/5}$ power law. When the accretion 
rate $\dot{m}$ is  large the change of the properties of the flows
is considerable in the ratios. As long as the canonical 
model parameters remain unchanged, the ADAF models with large $p$, 
which represent the convection, seem to be ruled out by 
the observational data.
In Figure 2, the core radio luminosity is  shown as a 
function of the SMBH mass, with SMBH candidates whose 
mass estimates are available.  From top to bottom, 
for each set of lines, the mass accretion rate 
corresponds to  $10^{-2}$ and $10^{-4}$. The open circles 
indicate the observed  core radio luminosity
at 15 GHz, and the filled circles the converted core radio 
luminosity  from 5 GHz to 15 GHz, as in Figure 1.
Although the radio flux data is in high angular resolution, 
there appears a somewhat wide range of radio luminosities,
which is likely the result of radio jets or other components
of various frequencies rather than pure ADAF emission. 
In Figure 3, the X-ray luminosity for the 2-10 keV band 
is shown as a funciton of the SMBH mass.

Radio flux values at 15 GHz of Sgr A*, NGC 1068 are quoted 
from \citet{kor95,gal96}, 
NGC 3079, NGC 3628, NGC 4151, and NGC 4388 are quoted from \citet{car90},
NGC 3031 (M81), NGC 4258, NGC 4736 (M94), and NGC 5194 (M51) 
are quoted from \citet{tur94}, respectively.
Radio flux data at 5 GHz of  NGC 1365 and NGC 3310 are quoted from \citet{sai94}, 
and NGC 224 (M31), NGC 3377, NGC 4374 (M84), NGC 4486 (M87), 
and NGC 4594 (M104) are quoted from \citet{fvf98}, respectively.

X-ray fluxes (2-10 keV) of Sgr A*, NGC 224, NGC 1068, NGC 1365, NGC 3031,
NGC 3079, NGC 3310, NGC 3628, NGC 4151, NGC 4258, NGC 4374, NGC 4388, NGC 4594,
NGC 4736, and NGC 5194 are quoted from 
\citet{koy96,tri99,tur97,iyo97,ish96,cdm98}; \citet{ptk99};
\citet{ptk98,nan97,mak94,cm99}; Iwasawa et al. (1997), Forster et al. (1999);
\citet{ptk98,rwo99,ter98}, respectively.
Quoted X-ray fluxes of NGC 4486 and NGC 3377 should be considered as upper limits
(Reynolds et al. 1999; Pellegrini 1999). Note  that the observed
X-ray flux of NGC 3377 in  the energy band of 0.2 - 4 keV is converted to 
the value in the energy band of 2 - 10 keV, assuming 
a power-law with the canonical spectral index 
for Seyfert 1 galaxies, i.e., $\Gamma = 1.7$.
High resolution X-ray images have revealed  discrete X-ray sources for 
NGC 224 (Trinchieri et al. 1999), NGC 3031 (Ishisaki et al. 1996),
NGC 4736 (Roberts et al. 1999). 
For these sources we adopt core or bulge X-ray fluxes.

The mass estimates of Sgr A*, NGC 1068, NGC 3031 (M81), NGC 3079, NGC 4374 (M84), 
4594 (M104)  come from \citet{eck97,gre96,ho99,fer00,ho99,fer00}, respectively.
NGC 224 (M31) is one of strongest SMBH candidates because of 
the fast rotation and the large velocity dispersion \citep{rich90}.
The estimates of SMBH mass come from 
Dressler \& Richstone (1988), Bacon et al. (1994).
NGC 224 seems to harbour a double nucleus \citep{lau93,bac94}.
The mass estimate of NGC 3377  comes from
Kormendy \& Richstone (1995) and Magorrian et al. (1998).
Besides M32, NGC 3377 is the second elliptical galaxy with 
stellar dynamical evidence for SMBHs.
The mass of NGC 4151 have been measured by 
reverberation mapping method \citep{wan99,geb00b}.  
The mass estimate of NGC 4258 comes from \citet{miyo95}.
NGC 4258 belongs to LINERs (e.g., Osterbrock 1989), which is one of 
successful applications of ADAFs \citep{las96}.
The mass estimates of NGC 4486 (M87) come from 
Ford et al. (1994), Reynolds et al. (1996), Magorrian et al. (1998), and
Ho (1999). NGC 4486 is a radio-loud elliptical galaxy with an optical jet 
which is approximately perpendicular to a disk of ionized gas.

\section{DISCUSSION AND CONCLUSION}

One of important predictions of the ADAF model is the radio/X-ray 
luminosity relation which can be directly used to estimate central
SMBH masses. We demonstrate with an improved statistical 
significance that if ADAFs are confirmed by observations 
with a high angular resolution, 
the inherent radio/X-ray luminosity relation provides a direct 
estimate of the central SMBH mass.
Several nearby extra galaxies have consistent ratios of radio to X-ray 
luminosities with the ADAF predictions for an estimated mass of the 
central SMBH.
We plot ADAF model predicitons for the radio/X-ray luminosities
 in terms of the SMBH mass in Figure 1. 
The observations for several extra galactic objects are 
consistent with the  predictions of  ADAF models, as
 seen in Figures 2 and 3. 
Although the mass estimates of NGC 4374, NGC 4486, NGC 4594 
appear more consistent with the predictions with truncated 
ADAF models or windy ADAF models, there is no obvious 
signs of the truncation or the wind in ADAFs. It is, however, interesting
to note that the models with the convection are exclusive unless
the microphysics in the standard ADAF model has to be modified
significantly. 
 
The predicted mass of NGC 224 (M31) is smaller than the estimated mass
at least by two orders of magnitude.
NGC 224 has an extrememly low core radio luminosity, which
is even lower than predicted with the mass accretion rate that 
the X-ray luminosity implies on the basis  of an ADAF model. 
NGC 224 seems to harbor a double necleus \citep{lau93,bac94}.
It is unclear whether ADAF is viable under such circumstance.
Predicted masses of NGC 1068, NGC 3031 (M81), NGC 3079, NGC 4151 are
$\sim 10^8 $ to $\sim$ a few $10^9~ M_{\odot}$. On the other
hand, reported mass estimates of SMBHs of these galaxies are less than $\sim 10^7$.
Masses of these sources are over-estimated due to high radio
luminosities resulting from strong jet-related activities.
Moreover, uncertainties in X-ray flux measurements and intrinsic variations
could cause errors in our predictions.
NGC 1068 is famous in that it has a strong compact radio source,
which results in a much higher radio/X-ray luminosity ratio than 
other Seyferts \citep{gal96}. 
The radio core of NGC 3031 (M81) is highly variable on many timescale
\citep{ho99}. 
Moreover, NGC 3031 is substantially brighter in X-rays
relative to the UV than in luminous AGNs \citep{ho99},
and its core luminosity in the 2-10 keV band
varies by a factor of $\sim 2$ in a timescale of years \citep{ish96}.
The unresolved radio map of NGC 3079 at 15 GHz shows an intriguing
feature, which  appears just like a jet \citep{car90}. The high
radio luminosity may be due to an unresolved small-scale jet.
The radio morphology  of NGC 4151 suggests that the emitting material
has been energized from the nucleus and that the jet interacts with 
the ambient medium \citep{car90}.

\acknowledgments

IY is supported in part by the KRF grant No. 1998-001-D00365.

\clearpage

\figcaption[fig1.ps]{A plot of the ratio of the radio luminosity 
to 2-10 keV X-ray luminosity versus the X-ray luminosity. 
The sources are spatially resolved, and the core luminosities 
are adopted in order to avoid possible contaminations due to
other components such as jets. SMBH masses are denoted by the
log scale at the top of each curve.
See text for detailed discussions and references. \label{fig1}}

\figcaption[fig2.ps]{The core radio luminosity is
shown as a function of the SMBH mass, with SMBHs whose masses are 
available. Mass accretion rates are denoted by the log scale at
 each line. Lines correspond to same models 
as in Figure 1 and symbols represent same observational data points 
as in Figure 1.\label{fig2}}

\figcaption[fig3.ps]{Similar plot as Figure 2. 
But the X-ray luminosity for 
the 2-10 keV band is shown, with the mass-known SMBHs. 
Lines correspond to same models as in Figure 1. \label{fig3}}

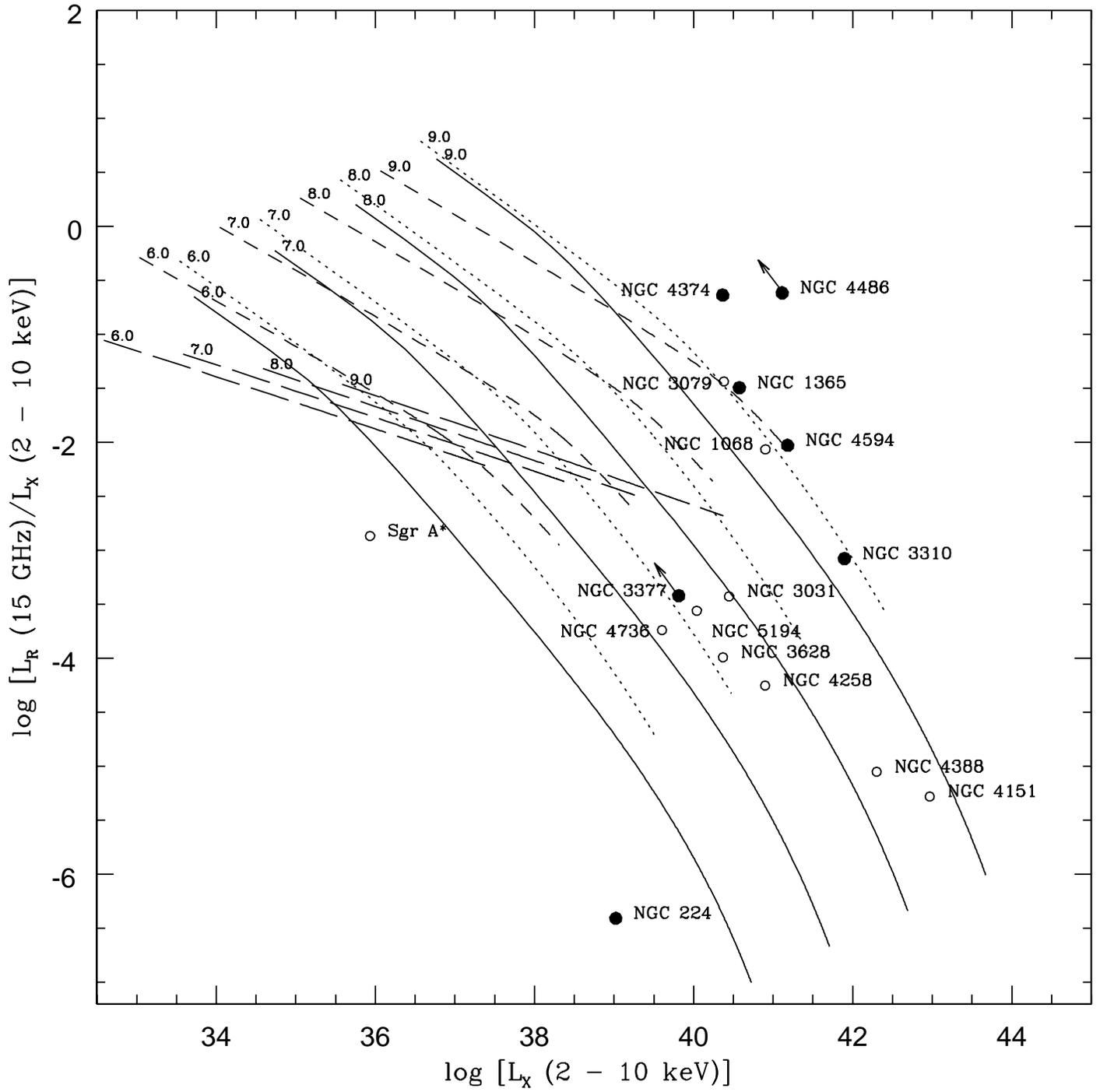
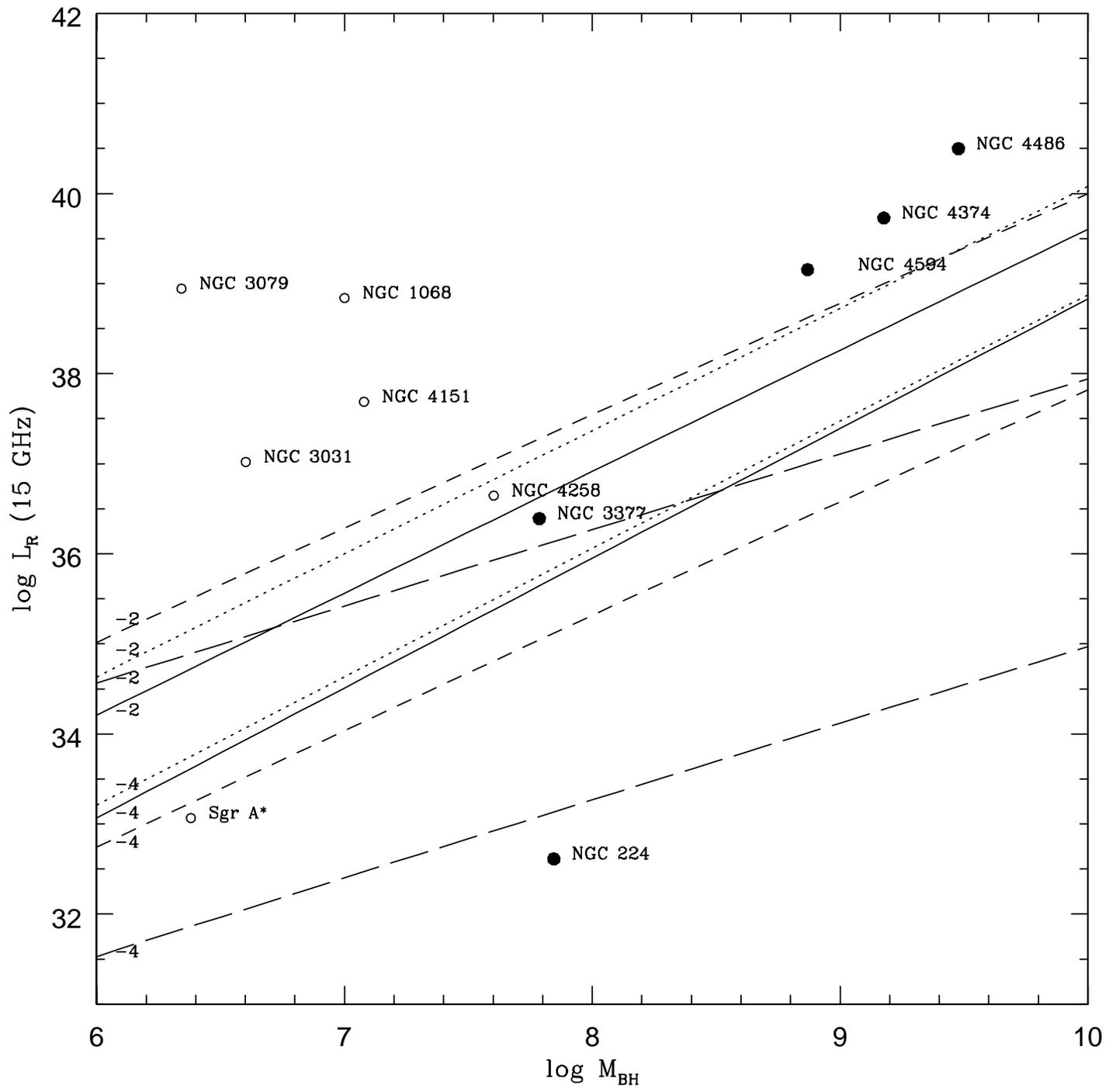
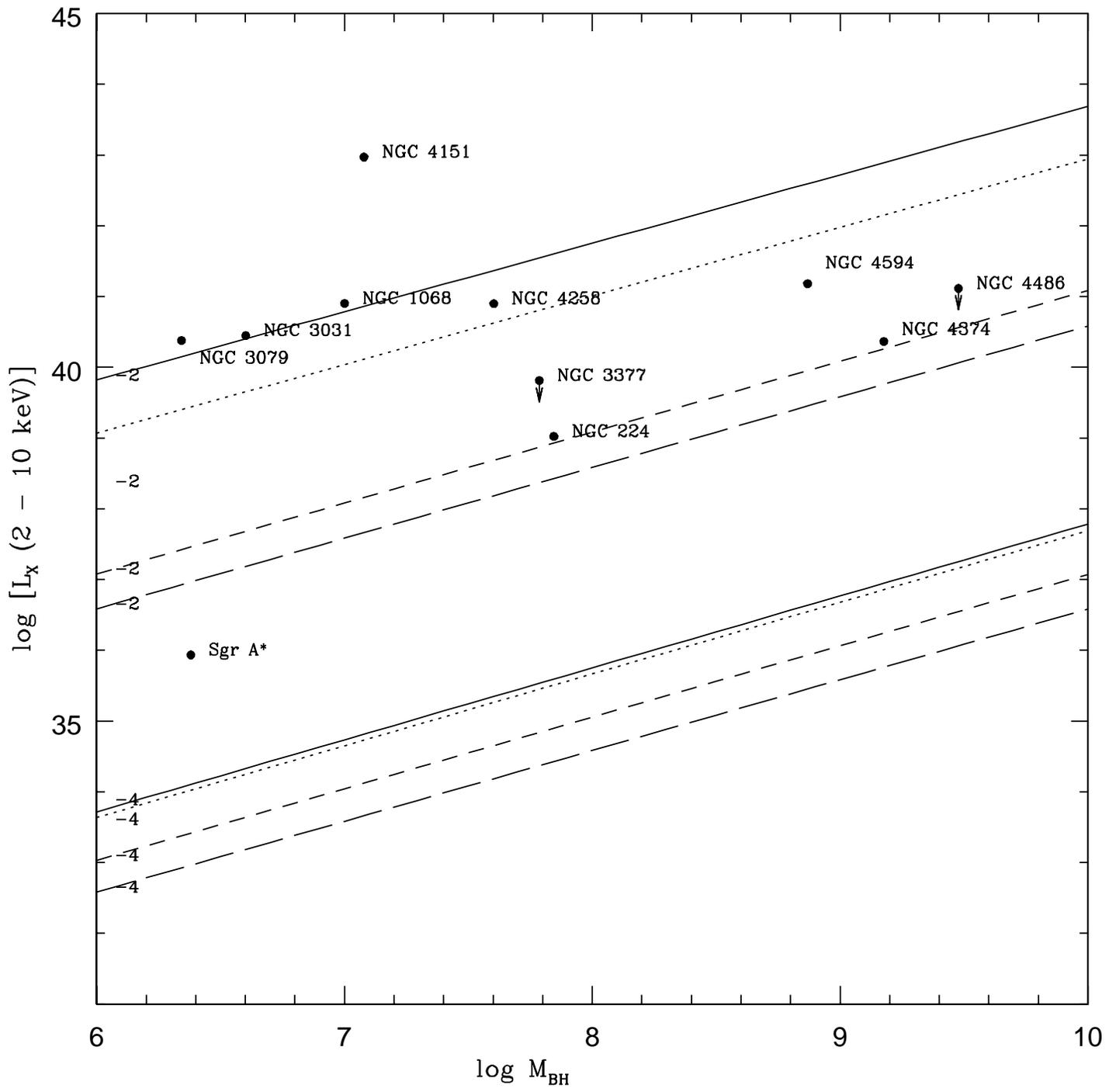
\end{document}